# Fool's gold? Developer dilemmas in a closed mobile application market platform

Joni Salminen, Jose Teixeira

Turku School of Economics
University of Turku, Finland
{Joni.Salminen,Jose.Teixeira}@utu.fi

**Abstract.** In this paper, we outline some potential conflicts that platform owners and software developers face in mobile application markets. Our arguments are based on comments captured in specialized online discussion forums, in which developers gather to share knowledge and experiences. The key findings indicate conflicts of interests, including 1) intra-platform competition, 2) discriminative promotion, 3) entry prevention, 4) restricted monetization, 5) restricted knowledge sharing, 6) substitution, and 7) strategic technology selection. Opportunistic platform owners may use their power to discriminate between third-part software developers. However, there are also potential strategic solutions that developers can apply; for example diversification (multi-homing), syndication and brand building.

## 1 Introduction

Although the rise of mobile application markets is a fairly recent phenomenon, using platforms[1] for distributing software, however, is not a new idea. For example, Linux has since long incorporated a centralized "market" for distributing software, while also introducing the one-click launch of applications. In addition to distribution[2], application-platforms enable developers to monetize their offerings and capture new customers. Therefore, developers can delegate to the platform owner critical business functions while concentrating on development – a benefit often appreciated by technology-focused founders.

Platforms are associated with network effects [1]: the more consumers a platform accumulates, the more feasible it is for developers to join, and vice versa. Because reaching a critical mass either side is costly, the platform owners require resources. Therefore, platforms compete fiercely against one another, and the winners achieve high returns. Due to the fast-growing mobile application industry, this paper is highly topical. Its developer perspective sheds light to how developers perceive platforms as partners. For platform owners interested in improving their attractiveness among developers, this information is crucial. As demonstrated by Nokia's failed effort with the Symbian/Meego platform, strategic choices in regards to the ecosystem matter to great extent. Although Nokia invested $250 million to the Symbian OS ownership, it was unable to increase the performance of the platform, thereby losing many developers and "killer apps" that attract end-users [2]. This forced the company to form a strategic alliance with Microsoft, simultaneously losing all future platform owner returns. Finally, awareness of potential conflicts of interests is useful for developers deciding whether or not to join a platform; or to form alliances to counterbalance the dominant position of the platform owner.

---

[1] 'Platform' is used in this paper to refer to commercial platforms, consisting of a platform owner and participants. A similar term is 'ecosystem', used in some fraction of literature; furthermore, the commercial perspective differentiates our approach from the fraction of IS research dealing with technological structure / infrastructure of a platform. Hence, our use of the term is close to 'marketplace'.

[2] Including transaction fees, bandwith costs, etc.

## 2 Mobile Application Platforms: A Review

Pioneered by Nokia's Ovi Store, the history of mobile application marketplaces consists of several market makers. This study will focus on three major platforms, namely Apple's App Store, Google Play, and Microsoft's Windows Marketplace – inarguably, there are more, such as RIM's BlackBerry and Samsung's Bada, but the market is likely to stabilize in oligopoly. For a detailed history of mobile application market development, see [3]. Table 1 exhibits details of the three major mobile application platforms.

**Table 1.** Comparison of the three major mobile application platforms

|  | App store | Google play | Windows Marketplace |
|---|---|---|---|
| Launch year | 2008 | 2012 | 2009 |
| Application base | 650K | 450K | 70K |
| Developer base | 215K | N/A | N/A |
| User base (devices) | 218M | 396M | 11M |
| Downloads | 26B | 14B | N/A |
| Revenue-sharing (%) | 70-30 | 70-30 | 70-30 |
| Entry fees | $99/year | one-time $25 | one-time $99 |
| In-platform marketing tools | Rankings + Featured apps | Rankings + Staff picks | Rankings + Spotlight |

A recent survey [4] discovered that developers find the installed user base the most important factor of choosing a platform. Interestingly, this surpassed revenue potential which was ranked only as fourth important factor. Furthermore, low development costs were ranked above revenue potential. In terms of economic rationality, developers could be expected to report the inverse; that the revenue potential is more than investments, especially considering that they are relatively low and differ a lot less than revenue potential. The explanation could be found in risk/loss aversion, or lack of even minimal funds. Further, it was discovered that developers would appreciate promotional tools which now, as revealed in the last row of Table 1, are largely missing.

Not only by their popularity among demand-side participants, have platforms also varied by their degree of openness. According to [5], a platform is open when its 1) development, commercialization, or use is not restricted, and 2) restrictions which exist to secure conformity (e.g. technical standards) are non-discriminatory, meaning that they apply uniformly to all participants, and reasonable. According to this strict definition, none of the three mobile application markets are open, but their degree of openness varies. More specifically, Table 2 examines their degree of openness by replicating the descriptive analysis done by [5]. Whereas they were focused on various operating systems, namely Linux, Windows, Macintosh, and iPhone, our analysis is on the three mobile application platforms.

**Table 2.** Elements of openness and closeness in mobile application platforms

|  | App store | Google play | Windows marketplac |
|---|---|---|---|
| Demand-side user (end user) | Closed | Closed | Closed |
| Supply-side user (developer) | Open | Open | Open |
| Platform provider (Hardware/OS) | Closed | Open | Open |
| Platform sponsor (Design & IPR) | Closed | Open | Closed |

For example, entry to applications market is controlled strongly by Apple, whereas Google is more liberal in accepting new applications. Further, access to the mobile device's capabilities and/or features may be restricted; development platform cannot be augmented e.g. by developing extra tools, and so on. Demand-side is closed because platforms are mutually exclusive and no interoperability exists; one cannot use Android applications with iPhone, but is "locked in". Access to a platform is granted given that the consumer purchases a certified device. This requirement is the strictest in the case of App Store, which not only is accessible only through Apple-manufactured devices but requires the user to commit his/her credit card number prior to participating. Developer lock-in is less severe (open supply-side); although applications developed for a particular platform cannot be directly translated to another platform, some part of their code base, visual appearances and programming logic can be reused across platforms. Further, participating requires purchasing a license may present a barrier for some developers, even though the cost is substantially smaller than for closed systems in many other industries (e.g. video game developer licenses). Developer tools can be found free of cost for all the major platforms, indicating that their technological basis is relatively open.

More specifically, a closed application platform is one owned and governed by a single party of authority which sets the rules for accepting, distributing and ranking applications entered in its system by third party developers. Closed platforms are often associated with proprietary, legally protected technologies [6]. Therefore, a closed platform dominates in its relationship to' developers; in contrast, open platforms set their rules based on consensus, e.g. open standards of technology, and can be forked[3] at any time.

In conclusion, modern mobile application platforms are somewhat a mixture of open and closed elements. Control is important for platforms, as history teaches prominent examples of failure in platform management. For example, IBM was the inventor of the PC market, but lost control to Microsoft – nowadays, Windows is the dominant PC operating system, whereas IBM has shifted its focus to consulting services. Another example is Atari, which was a dominant game console manufacturer at early 1980s. The company was unable to prevent the entry of opportunistic developers, flooding the market with low-quality games. This resulted in a "lemon's market" in which consumers could not distinguish good titles from bad ones, and eventually abandoned the platform altogether [7][4]). Descriptive in these cases is that once either side of a dual-market platform exits in sufficient quantities, the other side is quick to follow[5].

## 3 Potential conflicts: platform owner vs. developers

In this section, we outline some potential conflicts of interest between developers and platform owners. The arguments are backed to some extent by the voice of the developer community. More specifically, the conflicts are based on observations made by following discussion forums in which developers share experience and knowledge to topics relating to platform markets. Such a source of information is rich in nature and can quickly reveal the general attitudes and pain points relating to the relationship between developers and platform owners. We acknowledged that the developer view is inherently biased and attempted to choose only such vocations of opinion that included facts or plausible experiences while excluding speculative and hostile arguments which displayed strong negative attitude towards the platform owner and therefore risk being more biased than others.

The exhibits in Table 3 were gathered from several developer discussion forums. Clearly, they indicate that not all developers are satisfied by platform owners' use of power.

---

[3]  Forking refers to creating a new project based on the original code base
[4]  The original concept of "lemon's market" originates from Akerlof (1970).
[5]  Note the previous concern supply-side restrictions; demand-side restrictions exist as well but are not in the scope of this paper.

**Table 3.** Points of conflict / abuse of dominant position

| Point of conflict | Evidence from developer communities should be developers own month |
|---|---|
| 1) Intra-platform competition | *"for every serious application developer leaving the Mac App Store, there are 50 App developers moving-in"* |
| 2) Discriminative promotion | *"Recently, Apple changed their App Store ranking algorithm to stop ranking apps by download counts and instead use something else"* … *"Apple has also started rejecting pay-per-install apps ('freemium' apps that request the user to install companion apps to earn in-game currency)"* |
| 3) Entry prevention | *" I have made 6 builds trying to make MPlayerX pass Apple's review and I have explained why some privileges are so important for MPlayerX to achieve this and that features, But the answer is NO, NO, NO, NO, NO and NO."* <br> *"MPlayerX will lose so many features if it adopted Sandboxing, it could not load the subtitle automatically, it could not play the next episode for you automatically, it could not save the snapshots to the place where you want to, etc. Without those features, MPlayerX were just another lame Quicktime X, which I could not accept".* |
| 4) Restricted monetization models | *If Apple were a car company, they would already own the cars (the devices), and the roads (iTunes), and now they want a cut of the gasoline (content) money. They could certainly try, but it's up to you whether you will put up with it.* |
| 5) Restricted knowledge sharing | *"The problem is that the app store does not provide any stats on the number of downloads. You can get a ranking, but that tells you nothing about download numbers"* … *"no one (except Apple) knows exactly how the ranking algorithms work"* |
| 6) Substitution | *"Apple seems to be preemptively "Sherlock-ing" their most prosperous Mac Devs about one OS version BEFORE Apple copies them"* … *"Apple has a history of driving away developers by incorporating their ideas into the bundled apps. Not many developers though... only those of really well thought out OS enhancements"* |
| 7) Strategic technology selection | *Why do you think Windows and Mac are moving to an app store model? Sure, there's profit, but there wouldn't be any profit if people thought they were no more convenient than downloading from random websites.* <br> *" Isn't it obvious that platform manufacturers profit by limiting the access/content developers have to their systems?* |

**1) Intra-platform competition** – one of the prime motivations for joining a platform is that the platform owner adopts the role of marketer, promoting the application and enabling access to a pre-existing customer base. However, the observation is only superficial. As soon as a critical mass of suppliers (first movers) have entered the market, it becomes unfeasible for the platform owner to continue supporting all developers in their marketing. Instead, it opts for increasing rivalry among participants, thereby aiming to increase quality while allowing for natural selection of market mechanism to determine the winners. The supporting strategy is limited to select providers which are willing to pay or are otherwise critical in strategic interest of the platform ("killer apps"). Thus, joining a platform does not eliminate the need for marketing but, as a result of intra-platform competition, the entrant is forced to find ways of differentiation and prominence, controlled by the platform owner. Essentially, the ore developers compete, the more this supports the platform owner's goal to build high-quality offerings for consumers – for developers, obviously, fewer competitors would be better.

**2) Discriminative promotion** – it is impossible for a platform owner to provide fair treatment in rankings due to the high number of entrants. Even by diving them into several categories, the platform owner is unable to guarantee a fair exposure to customers. The notable exceptions include positive discrimination of novel apps and other measures that secure constant and equal rotation of featured apps[6]. The uneven popularity of apps is not only reliant on their presentation in featured positions but their rise to that position results from other factors; then under the protection of the platform owner the initial success becomes a virtuous cycle. In fact, the platform owner optimizes its own revenue by displaying most popular applications at most prominent positions while encouraging developers to compete against in marketing activities beyond the platform. The effect of promotion shows in "jumping the shark", whereupon temporary promotion brings a spike of new users, but once featured apps rotate, the growth will rapidly decline [8]. Therefore, it makes sense for the platform owner to promote by rotating apps in featured positions; this keeps many of them satisfied but allows none to maximize their prominence; which would be the goal of an individual developer.

**3) Entry prevention** – apart from favoritism in promotion, the platform owner's strategic interest can take the form of protectionism, preventing the entry of undesired applications. After a threshold of critical mass, the marginal increase in utility experienced by consumers does not increase substantially by additional applications (excluding "killer apps" which represent outliers). Therefore, the platform owner may enforce his selection power, labeled e.g. under quality control[7], to prevent entry e.g. from such applications that are perceived too similar to incumbent category leaders (imitation). This type of protectionism can also be seen in censorship of applications that violate the terms of service by the platform owner, and favoring ones that have gained traction as a function of social diffusion. Various analyses show that platform owners can increase their profit by decreasing competing among participants [9]. This is especially valid for saturated marketplaces, in which marginal value of newcomers is diminishing.

---

[6] However, if the rate of new apps introduces to each category is sufficiently high, it may be unpractical to exercise this type of discrimination – consider a user unable to find a popular application because the leaderboard changes each time.

[7] Quality control is only a problem when the platform owner shadows unfair filtering in the normal process of quality policy.

**4) Restricted monetization** - it is common that developers apply a myriad of monetization models to maximize their profits. These may include, apart from direct application purchases solicited by the platform owner, in-app purchases (e.g. virtual goods), subscriptions, paid content and such, see e.g. [10]. However, in the case of monetization beyond the platform, the platform owner is cut off from the revenue capture. Because the platform owner is even in this case dealing with user acquisition and distribution tasks allocated to it, it feels violated when revenue capturing takes place outside. Therefore, it may attempt to prevent such activity, for example by disallowing applications with links to external payment providers. Keeping all economic activity within the platform is in the strict interest of the platform owner, as its returns are directly proportional to purchased applications. This naturally limits the available monetization models for developers, although there are exceptions. Facebook, for example, is running its App Center through its mobile applications; using its own technology to disseminate and monetize offerings. However, this is possible only for rare cases with strong demand-side position, so that platform owners cannot exclude them without harming the platform as a whole. In other cases, the platform owner will control the ability of participants to monetize their offerings, while developers would rather see unlimited options for monetization[8].

**5) Restricted knowledge sharing** – the platform owner may be tempted to hide information such as download statistics for competitive reasons. It may also prohibit in its terms of service (TOS) the sharing of such information, and censor this and other types of information sharing by developers in its information outlets, such as the official discussion forum. Such an activity is exhibited by cases of deleting unfavorable messages [example], as well as general refusal to share aggregate data on download and usage statistics. Naturally, platform owners use this data to improve their position in inter-platform competition – the alternative cost, from developer's perspective, is the loss of transparency.

**6) Substitution** – substitution may take place through acquisition or rivalry. As noted by [5], absorbing an application can provide demand-side advantages. While individual developers receive high payoffs in most acquisitions, rivals of such an application are likely to face restrictions by the platform owner as it is likely to favor its own application. Such a move would effectively remove competition in this specific sub-vertical of the application market. Another form of substitution is rivalry; for example, the platform owner can absorb ideas into its own offering. In inter-platform competition, absorbing can be seen as expanding core software at the expense of developers contributing to the intra-platform category. Substitution hazards do not apply when the platform owner is explicitly not involved in the market in other role than the owner, which in theory guarantees a fair treatment of the joined developers. In other cases, it is rational for the platform owner to absorb applications of functionalities that are strategic in inter-platform competition, regardless of harming an individual developer in doing so.

---

[8] However, the platform owner will support the diversification of monetization, as long as it takes place within the platform

**7) Strategic technology selection** – developers are interested in leveraging the latest technology to provide the best experience for consumers. However, platform owners are tempted to not adopt technologies that require substantial investments, are not matched by competitors, or demise their own technological infrastructure. Eventually though, platform owners are forced to adopt technologies due to their rivalry; however, delaying such a decision until competitors make similar investments is a strategically feasible option. Obviously, the welfare loss of delayed adoption will be paid by consumers of applications. While most developers would enjoy the release of new technology, it may at times be against the platform owner's strategic agenda.

## 4 Developers solutions

In the following table, we outline some prominent solutions developers can opt for to reduce their risk and increase their bargaining positions in regards to the platform owner.

table 4

**A) Custom development** - the developer may aim to enjoy the application hype indirectly by leveraging his skill to develop applications for others in exchange for payment, in effect leveraging hype externalities. In this model, the client assumes the risk of failure while returns for developer are guaranteed. This effectively solves the problem of limited monetization opportunities because the developer is free to choose his monetization model. Further, to enhance his position, the developer can pursue contractual revenue sharing, in which case he would exchange a part of his compensation for eventual scale returns of the client's 'killer app'. Therefore, according to his risk tolerance the developer can apply a mixed revenue model. By capturing hype externalities, the developer is also able to hone his skills, regardless of the technology stack offered by the platform owner. This is beneficial in cases of strategic technology selection. The natural consequence is that it tends to increase the amount of information pertaining to the focal topic. In other words, hype effects can be used by the developer to benefit from knowledge sharing and to increase their skills.

**B) Diversification (multi-homing) -** A common and fairly straight-forward strategy is diversification, also sometimes called multi-homing or multi-platform strategy, in which the developer offers applications in several platforms, thereby reducing his dependence on a single source of revenue, and also extending his repertoire for monetizing. In a case of rejection from one platform, the developer is free to apply elsewhere. The negative effects include additional labor required to adapt the software to several platforms, and duplicating potential within platform marketing efforts. However, beyond-platform marketing is synergistic as long as only the platform (not application) changes, and there is learning overlap in developing for several mobile platforms. The major disadvantage of the method is of course the need for investing into several development projects – although there are potential development synergies between platforms, developers may consider possible differences in user bases and other characteristics of the market (e.g. purchase willingness; ratio of paid and non-paid applications, etc.).Diversification also gives access to various technological infrastructures, bypassing platform owner's strategic technology selection. Further, it protects against substitution; when substitution takes place, the developer shifts his focus over to others. However, diversification will not solve intra-platform competition because in each platform he enters, the developer finds new rivals. Large developers are able to diversify titles to increase the likelihood of creating hits, while a small developer has to succeed with fewer attempts.

**C) Brand building** - strong branding increases the potential of becoming a 'killer app', followed by additional support from the platform owner. Brands are an effective means to distinguish from the competition, and they may be under-exploited by technologically oriented founders. A recent example of a successful branding strategy includes Rovio's Angry Birds that has spun off to a tremendous number of product categories not related to technology at all. A strong brand can act as a shield against substitution; it is harder to replicate than functionalities – allowing, and in fact encouraging the birth of 'killer apps' is a stronger goal for the platform owner than recreating or absorbing them1. Thus, branding not only offers means to differentiate in intra-platform competition but also to pre-empt substitution threats to a major degree. In addition, the platform owner is unlikely to exclude entrants from branded applications[9]. When brand building efforts take place outside the platform, a developer is able to circumvent the platform owner's discriminative marketing.

**D) Cross-promotion** - an example of within-platform marketing, cross-promotion typically operate in "tit-for-tat" fashion; in other words, one referred user from application A to application B earns one back in exchange. Through technological mediation such activities can be coordinated efficiently, as proven by the Facebook application 'Applifier' which reached over 55 million developer-users within its first ten months. Indeed, cross-promotion is a means of turning around the intra-platform competition; instead of being eaten away by each other, developers may share users; although it is vulnerable to the same power law dynamics than regular application rankings/listings[10], it presents an alternative to the market mechanism provided by the platform owner, reducing the negative effects of discriminative marketing.

**E) Syndication** - finally, developers may discriminate abusive platform owners and promote their misdeeds to other developers. Because the platform owner is dependent on developers as a collective unit, the response of abuse should also take a collective form. E.g. protest, bans, and such are means that can be used to generate negative press for the platform owner and pressure it to increase transparency and fair policies. Thereby, harnessing 'group power' in an organized means would produce better outcomes than atomistic complaints. For example, developer communities are a natural means for online re-grouping, and a substantial part of knowledge sharing takes place inside them. By syndicating, developers are also able to communicate claims regarding to the future development of the platform's technology stack, thereby influencing the strategic behavior of the platform owner. If the platform owner remains ignorant to organized requests by the developer community, it risks losing popularity – therefore, syndication signals credible threats. Syndication can also protect from intra-platform competition; instead of competing against each other, developers may share resources and skills according to formalized agreements. At the same time, the platform owner may find it more difficult to substitute syndication as oppose to single developer.

---

[9] However, it will exclude those aiming to leverage the platform owner's brand, e.g. iTunz and similar variations are not allowed in Apple's App Store.

[10] The more is share, the more is received; this favors a winner-take-all structures, as the most popular applications are accounting for most shared traffic. Some means to normalize distribution include thematic inclusion (in which only thematically relevant applications are shown to the user); and limited entry (naturally not all developers are joining cross-promotion systems, which gives those who join some advantage over them).

**F) Creating network effects** – the structure of application markets supports network effects e.g. [7]. In one sided markets, network effects relate to growing the homogeneous user-base: classical examples include railroads and telephone lines: the more they have coverage, the more it makes sense to use them in various situations. In dual sided markets, the user groups are heterogeneous and typically complementing one another, e.g. men and women in a dating application, or buyers and sellers in a marketplace. Harnessing network effects enables an application to bypass the within-platform marketing; typically they spread outside the platform as a function of word-of-mouth or other forms of peer-marketing; independent from the existence of the platform as a marketing channel, but dependent on it as a distribution channel. Such a situation effectively solves the conflict of interests: the platform owner is not required to use resources for marketing, the developer is not dependent on the ranking effects within a platform, and both earn revenue according to the set revenue-sharing scheme. Further, network effects protect against substitution – the platform owner may more easily replicate the functionality of an application than its user base. This can be seen in many cases of acquisitions, in which a fairly simple application is acquired due to its enormous user base[11].Network effects are a natural means of solving intra-platform competition; since they reach beyond the platform, the developer is able to draw support from outside, as opposed to being vulnerable to the platform owner's discriminative promotion tactics.

## Conclusions and discussion

In conclusion, the conflicts are the result of power imbalance in distribution and marketing of applications. Discriminating between applications is a strategy that the platform owner can use to maximize its revenues. Further, by creating a winner-take-all structure in its categories, it can enhance the reference point effect, i.e. creation of killer apps that media is more likely to cover and newcomers are likely to see as their role models. In an increasingly competitive application market, the lack of transparency reduces developers' ability to make just decisions. For example, the role of marketing agencies in increasing the ranking of applications either legitimately (e.g. negotiating with Apple to increase ranking to featured position) or by fraudulent measures (e.g. fake ratings, downloads) not only place developers in unequal positions but disrupt the market mechanism in determining which applications receive most prominence. These issues are no different from classical economic arguments against non-centralized decision making e.g. [11]; thereby revealing the rating mechanisms, and if necessary altering them closer to market-driven demand would be a reassuring signal from the platform owners to alleviate the uncertainty faced by developers. In any case, developers are sensitive to structural changes by the platform owner; for example, when Facebook changed its ranking algorithm for Application Directory, to prefer usage over the number of downloads, the change lead developers to favor greater interactivity in their apps [7].

---

[11]   Such would be e.g. the acquisition of Instagram by Facebook.

Finally, the observation also formulates a strategic guideline for developers looking to monetize the growing app markets: finding and entering sub-categories without a leader can produce escalating rents intensified by the platform owner's protection. However, as the platform market matures, so do various sub-categories, and it becomes increasingly difficult to find new ones. There are strong basis to argue for "fool's gold" phenomenon when entering in a saturated category with little ability to differentiate against category leaders. In conclusion, we have presented several potential conflict areas between platform owners and developers, as well as offered some prominent solutions as a strategic course of action for developers. It must be noted that both parties act according to their own interest, and the conflict of interests is therefore a natural outcome. Under this premise, the platform owner is interested in protection of its business interests while the developer wishes to maximize his share of revenue in the collaboration consisting of development, distribution and monetizing applications. In a mixture of economic and non-economic motives, however, gains of purely profit-oriented agents are somewhat diluted – this happens because consumers choose the cheaper option which satisfies their minimum requirements [12], while some fraction of developers is more interested in other effects of popularity rather than revenue; yet competing in the same platform.